# Influenza-associated mortality for circulatory and respiratory causes during the 2013-2014 through the 2018-2019 influenza seasons in Russia

Edward Goldstein[1,*]

1. Center for Communicable Disease Dynamics, Department of Epidemiology, Harvard TH Chan School of Public Health, Boston, MA 02115 USA

*.  Email: egoldste@hsph.harvard.edu

**Abstract**

Background: Information on influenza-associated mortality in Russia is limited.

Methods: Using previously developed methodology (Goldstein et al., Epidemiology 2012), we regressed the monthly rates of mortality for respiratory causes, as well as circulatory causes linearly against the monthly proxies for the incidence of influenza A/H3N2, A/H1N1 and B (obtained using data from the Smorodintsev Research Institute of Influenza (RII)), adjusting for the baseline rates of mortality not associated with influenza circulation and temporal trends.

Results: For the 2013/14 through the 2018/19 seasons, influenza circulation was associated with an average annual 17636 (95% CI (9482,25790)) deaths for circulatory causes and 4179 (3250,5109) deaths for respiratory causes, with the largest number of deaths (32298 (18071,46525) for circulatory causes and 6689 (5019,8359) for respiratory causes) estimated during the 2014/15 influenza season. The biggest contributor to both circulatory and respiratory influenza-associated deaths was influenza A/H3N2, followed by influenza B and A/H1N1. Compared to the 2013/14 through the 2015/16 seasons, during the 2016/17 through the 2018/19 seasons (when levels of influenza vaccination were significantly higher), the volume of influenza-associated mortality declined by about 16.1%, or 3809 annual respiratory and circulatory deaths.

Conclusions: Influenza circulation is associated with a substantial mortality burden in Russia, particularly for circulatory deaths, with some reduction observed following the major increase in vaccination coverage. Those results support the potential utility of further extending the levels of influenza vaccination, the use of quadrivalent influenza vaccines, and extra efforts for protecting individuals with circulatory disease in Russia, including vaccination and the use of antiviral medications.

**Introduction**

Annual epidemics associated with the circulation of influenza A/H3N2, A/H1N1 and B take place in Russia [1,2]. Such epidemics are known to result in a substantial toll of mortality in other countries in the Northern Hemisphere, e.g. [3-9], including mortality for circulatory causes [3,4], with influenza additionally known to be associated with a variety of cardiovascular manifestations [10,11]. At the same time, there is limited information on the burden of influenza-associated mortality in Russia. Such information is largely related to deaths with influenza in the diagnosis, e.g. [2,12-15]. However, cases with influenza in the diagnosis are known to represent only a small fraction of all influenza-associated deaths. For example, it was found in [11] that laboratory-confirmed influenza infection increases the risk of an acute myocardial infarction by a factor of about 6. However, influenza infection is detected/diagnosed extremely rarely for such excess myocardial infarctions and associated deaths. Additionally, cases with influenza in the diagnosis may not be representative of the relative contribution of the major influenza (sub)types (A/H3N2, A/H1N1 and B) to mortality. Indeed, the age distribution of diagnosed deaths with influenza A/H1N1 is significantly younger compared to diagnosed deaths with influenza A/H3N2 and B in Russia, e.g. [13]. This suggests that deaths associated with A/H1N1 infections are much more likely to be diagnosed compared to deaths associated with A/H3N2 and influenza B infections. Indeed, even during the 2014/15 season when influenza A/H1N1 circulation was minimal, influenza A/H1N1 was found to represent the largest contribution to diagnosed influenza-related deaths in Russia [14,15]. All of this suggests that better understanding is needed of the overall volume of influenza-related deaths in Russia, as well as of the relative contribution of each of the major influenza (sub)types to mortality, including changes in the above following the significant increase in influenza vaccination coverage starting the 2015/16 season.

In our earlier work [3,4,16], we introduced a new method for estimating the burden of severe outcomes associated with influenza and the respiratory syncytial virus (RSV), designed to address several limitations of some of the previously employed inference models. An important feature of that approach is the use of RSV and influenza A/H3N2, A/H1N1, and B incidence proxies that are expected to be linearly related to the population incidence of those viruses; such proxies for influenza (sub)types combine data on medical consultations for influenza-like illness (ILI) with data on the testing of respiratory specimens from symptomatic individuals [3,4,16]. We've applied those incidence proxies to estimate the rates of influenza-associated mortality stratified by age/cause of death in the US [3,4], with the corresponding method being later adopted for the estimation of influenza-associated mortality in other countries as well, e.g. [6-9]. In this paper, we derive the analogous proxies for the weekly/monthly incidence of influenza A/H3N2, A/H1N1, and B in Russia based on the surveillance data from the Smorodintsev Research Institute of Influenza (RII) [17,18] that includes information on medical consultations for influenza/ARI symptoms, data on testing of respiratory specimensfor the different influenza (sub)types, as well as data on the antigenic characterization of the circulating influenza strains. We then apply the inference model from [3,4,16] to relate those incidence proxies to the monthly data from the Russian Federal State Statistics Service (Rosstat) [19,20] on the rates of death for respiratory causes and circulatory causes to estimate the corresponding mortality rates associated with the circulation of the major influenza (sub)types during the 2013/14 through the 2018/19 influenza seasons in Russia.

**Methods**

*Data*

Monthly data on mortality for respiratory and circulatory causes of in Russia were obtained from [19]. Monthly mortality counts for respiratory and circulatory deaths were then converted to monthly rates of mortality for those causes per 100,000 individuals using population data from Rosstat [20]. Influenza is known to be associated with mortality for other causes as well, including neoplasms and metabolic diseases [3,4]; however finer (e.g. weekly) mortality data are needed to ascertain the signals from influenza circulation in the corresponding time series.

Weekly data on the rates of ILI/ARI consultation per 10,000 individuals in Russia are available from [17] (where they are called the influenza/ARI rates). Data on the weekly percent of respiratory specimens from symptomatic individuals that were RT-PCR positive for influenza A/H1N1, A/H3N2 and influenza B are available from [18].

*Incidence proxies*

Only a fraction of individuals presenting with influenza/ARI symptoms are infected with influenza. We multiplied the weekly rates of influenza/ARI consultation per 10,000 individuals [17] by the weekly percentages of respiratory specimens from symptomatic individuals that were RT-PCR positive for each of influenza A/H1N1, A/H3N2 and B [18] to estimate the weekly incidence proxies for each of the corresponding influenza (sub)types:

*Weekly influenza (sub)type incidence proxy =* (1)

*= Rate of consultations for influenza/ARI * % All respiratory specimens that were positive for the (sub)type*

As noted in [3], those proxies are expected to be *proportional* to the weekly population incidence for the each of the major influenza (sub)types (hence the name "proxy"). Monthly incidence proxies for influenza A/H1N1, A/H3N2 and B were obtained as the weighted average of the weekly incidence proxies for those weeks that overlapped with a given month; specifically, for each influenza (sub)type and month, the incidence proxy for each week was multiplied by the number of days in that week that were part of the corresponding month (e.g. 7 if the week was entirely within that month), then the results were summed over the different weeks and divided by the number of days in the corresponding month. To relate the incidence proxies for the major influenza (sub)types to monthly mortality rates we first shift the incidence proxies by one week forward to accommodate for the delay between infection and death [3,4], then use the shifted weekly incidence proxies to obtain the corresponding monthly incidence proxies as above.

The relation between an incidence proxy and the associated mortality may change over time. In particular, influenza B is characterized by the circulation of *B/Yamagata* and *B/Victoria* viruses. It is known that the age distribution for the *B/Yamagata* infections is notably older than for the *B/Victoria* infections [21,22]. Correspondingly, the relation between the incidence proxy (which reflects influenza incidence in the general population) and influenza-related mortality (which for influenza B largely reflects mortality in older individuals) may be quite different for influenza *B/Yamagata* compared to influenza *B/Victoria*. Influenza *B/Yamagata* dominated the influenza B circulation in Russia during the 2013/14, 2014/15, 2017/18 and the 2018/19 seasons, while influenza *B/Victoria* was dominant during the 2015/16 and the 2016/17 seasons [18]. Correspondingly, we split the incidence proxy for influenza B into two: one for the 2013/14, 2014/15, 2017/18 and the 2018/19 seasons (called the *B/Yamagata* proxy), and one for the 2015/16 and the 2016/17 seasons (called the *B/Victoria* proxy). During the 2014/15 and the 2015/16 influenza seasons, a drift variant of influenza A/H3N2 with reduced titers to the vaccine-type strains circulated in Russia [18]; moreover this variant was known to cause high levels of mortality in other countries [7]. During the other four seasons in our study, the circulating A/H3N2 strains were similar to the vaccine-type strains. Correspondingly, we split the A/H3N2 incidence proxy into two: one for the 2014/15 and the 2015/16 seasons (called the drift A/H3N2 proxy), and one for the 2013/14, 2106/17, 2017/18 and the 2018/19 season (called the matching A/H3N2 proxy). Finally, we know that for all six influenza seasons in the data, the majority of circulating influenza A/H1N1 viruses were similar to the vaccine-type viruses during the corresponding seasons. Figure 1 plots the monthly incidence proxies for influenza A/H1N1, A/H3N2 and B in Russia between

07/2013 and 07/2019 (73 months), with the proxies for influenza A/H3N2 and B further split into two as described in this paragraph.

*Inference method*

If $M(t)$ is the mortality rate for a given cause on month $t$ (with $t = 1$ for 07/2013), and $A/H3N2^{match}(t), A/H3N2^{drift}(t), A/H1N1(t), B/Victoria(t), B/Yamagata(t)$ are the incidence proxies for the different influenza (sub)types on month $t$ as described in the previous subsection, then the inference model in [3,4] suggests that

$$M(t) = \beta_0 + \beta_1 \cdot A/H3N2^{match}(t) + \beta_2 \cdot A/H3N2^{drift}(t) + \beta_3 \cdot A/H1N1(t) + \beta_4 \cdot B/Victoria(t) + \beta_5 \cdot B/Yamagata(t) + Baseline + Trend + Noise \qquad (2)$$

Here *Baseline* is the baseline rate of mortality not associated with influenza circulation that is *periodic* with yearly periodicity. We will model it as

$$Baseline(t) = \beta_5 \cdot \cos\left(\frac{2\pi t}{12}\right) + \beta_6 \cdot \sin\left(\frac{2\pi t}{12}\right) + \beta_7 \cdot \text{Jan}(t)$$

Here *Jan* is a variable equaling 1 for the month of January, 0 otherwise. The reason for including this variable is that the monthly (rather than annual) mortality data in [19] is operational, with some of the mortality not registered during a given calendar year being added to January on the next year [23]. Furthermore, the January effect (likely weather-related) on circulatory mortality in Russia was established in [24]. The (temporal) trend is modeled as a cubic polynomial in time (month). Finally, the sensitivity of the estimates of the contribution of influenza to mortality with regard to the model for the baseline is discussed in the Supporting Information for [3] (where limited sensitivity was found).

The Akaike Information Criterion (AIC) is adopted to select the variables (covariates) in the model in eq. 2 for each mortality cause, with the variable whose omission results in the largest decline in the AIC score being dropped at each step until no omissions of a variable decrease the AIC score. The model equations for respiratory mortality and circulatory mortality selected by the stepwise AIC criterion are as follows:

*Respiratory mortality*

$$M_{resp}(t) = \beta_0 + \beta_1 \cdot A/H3N2^{match}(t) + \beta_2 \cdot A/H3N2^{drift}(t) + \beta_3 \cdot A/H1N1(t) + \beta_4 \cdot B/Yamagata(t) + \beta_5 \cdot \sin\left(\frac{2\pi t}{12}\right) + \beta_6 \cdot \text{Jan}(t) + \beta_7 \cdot t + \beta_8 \cdot t^2 + \beta_9 \cdot t^3 \qquad (3)$$

*Circulatory mortality*

$$M_{circ}(t) = \beta_0 + \beta_1 \cdot A/H3N2^{match}(t) + \beta_2 \cdot A/H3N2^{drift}(t) + \beta_3 \cdot A/H1N1(t) + \beta_4 \cdot B/Yamagata(t) + \beta_5 \cdot \sin\left(\frac{2\pi t}{12}\right) + \beta_6 \cdot \text{Jan}(t) + \beta_7 \cdot t + \beta_8 \cdot t^2 \qquad (4)$$

We note that the $B/Victoria$ incidence proxy is not selected by the AIC for either model, whereas the other four incidence proxies are all present in each model.

**Results**

Figure 1 plots the monthly incidence proxies for influenza A/H1N1, A/H3N2 (both matching and drifted) and B (both Victoria and Yamagata) in Russia between 07/2013 and 07/2019. Figure 2 plots the daily rates of respiratory deaths and circulatory deaths by month between 07/2013 and 07/2019. We note that there is a good temporal correspondence between the "bumps" in mortality rates and peaks of influenza circulation, save for the sizeable wave of influenza *B/Victoria* in 2017. The lack of the corresponding visual effect of influenza *B/Victoria* circulation on mortality may be related to the young age distribution for influenza *B/Victoria* infections (Methods; [21,22]).

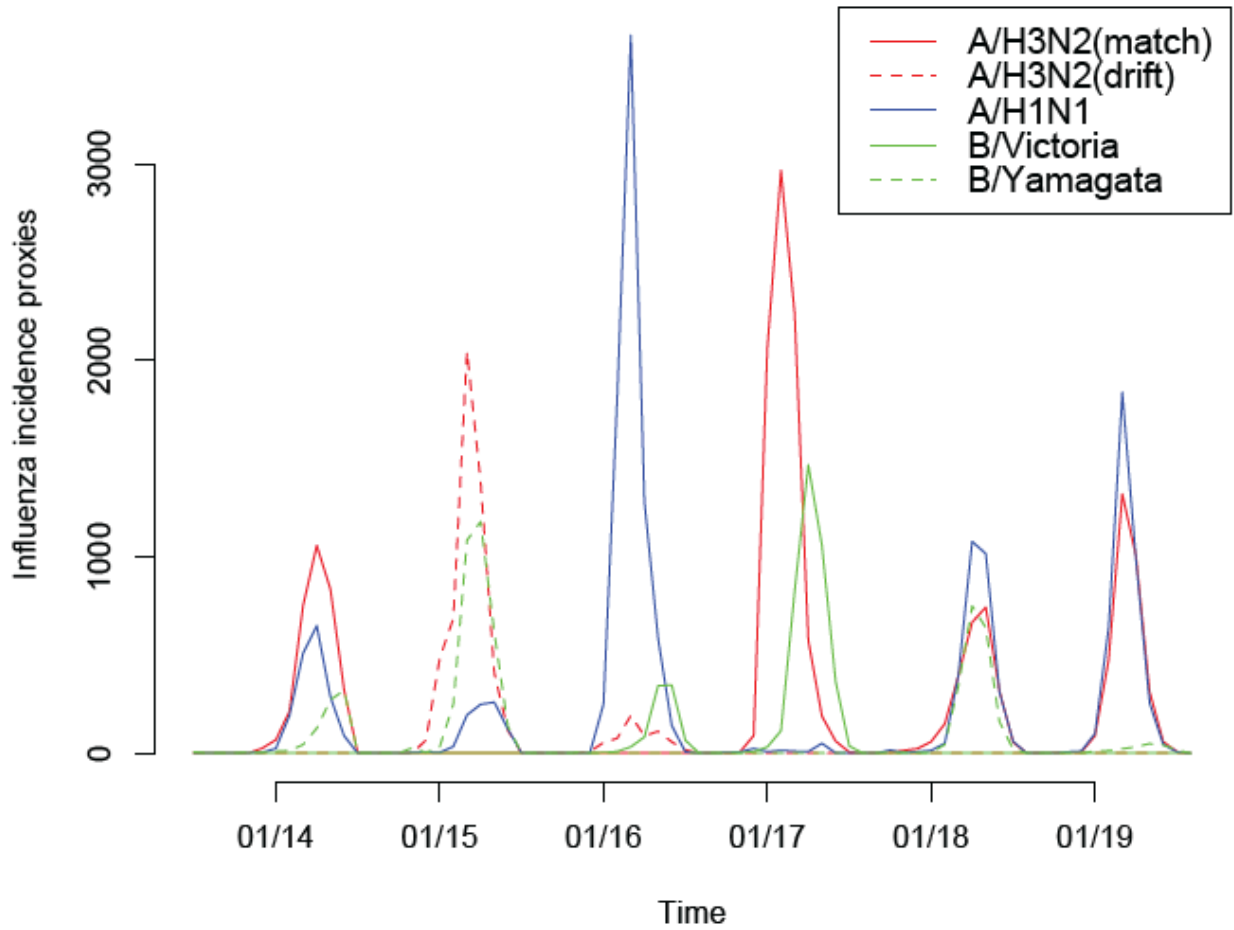

**Figure 1:** Monthly incidence proxies for influenza A/H3N2, A/H1N1 and B in Russia, 07/2013 through 07/2019.

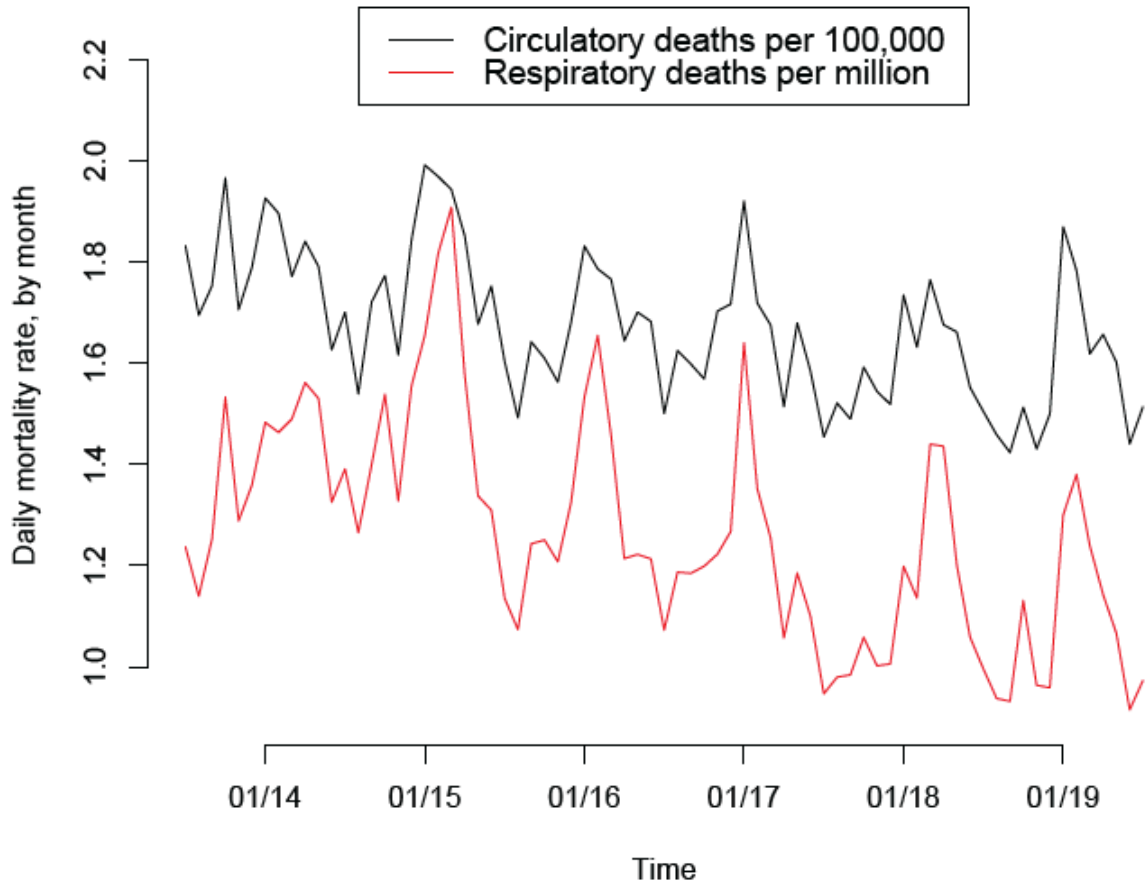

**Figure 2:** Rates of daily mortality for respiratory causes (per one million individuals in Russia) and circulatory causes (per 100,000 individuals) by month, 07/2013 through 07/2019.

Figure 3 plots the fits for the model for respiratory mortality given by eq. 3. Figure 4 plots the fits for the model for circulatory mortality given by eq. 4. The fits appear to be fairly temporally consistent, particularly for respiratory deaths, with the weekly contribution of influenza to mortality estimated as the difference between the red and the green curves in Figures 3 and 4. Those fits help support the basic structure of our model, namely that monthly rates of both respiratory and circulatory mortality in Russia are largely explained by influenza's contribution to the regular pattern mortality (baseline + trend).

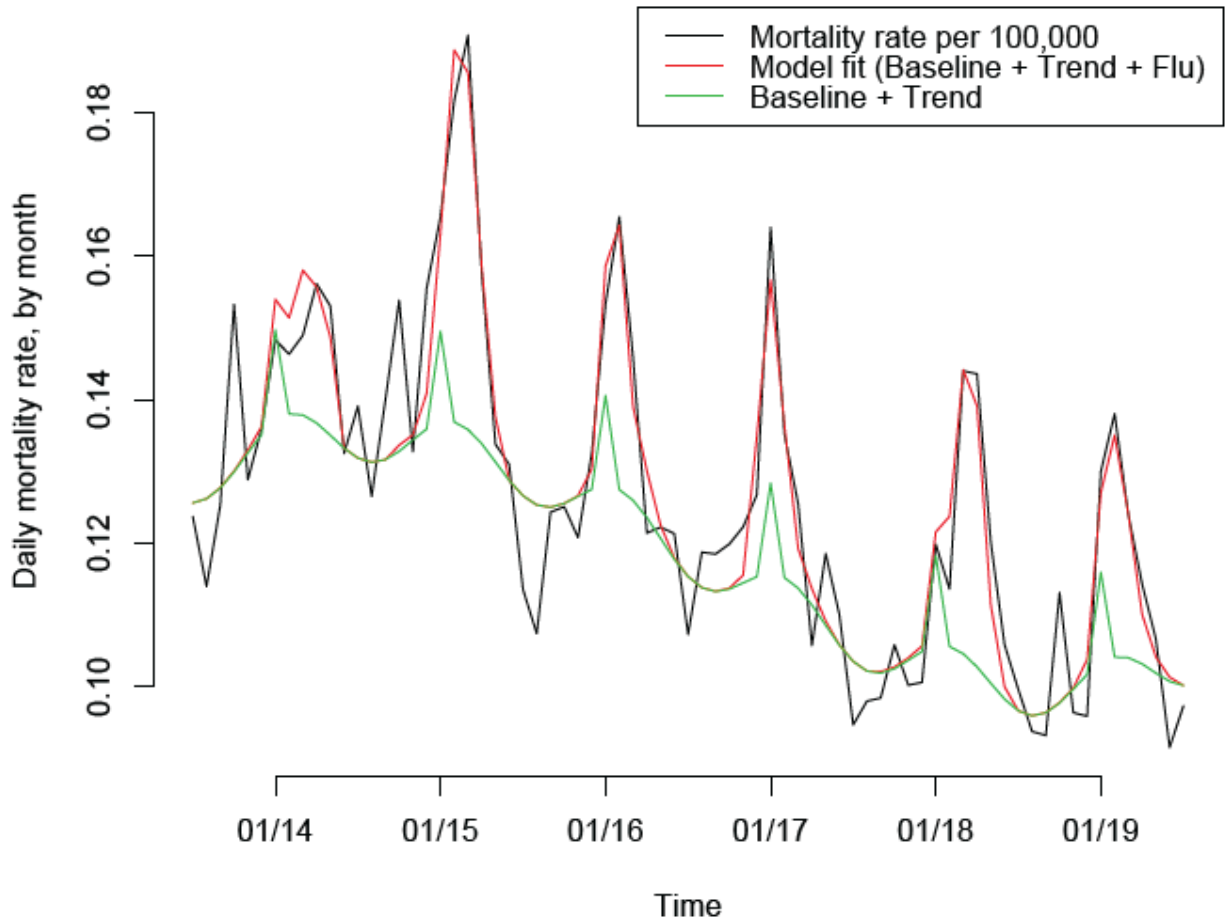

**Figure 3:** Model fit for the daily rates of respiratory deaths per 100,000 by month in Russia, 07/2013 though 07/2109. The contribution of influenza to respiratory mortality is represented by the difference between the red and the green curves.

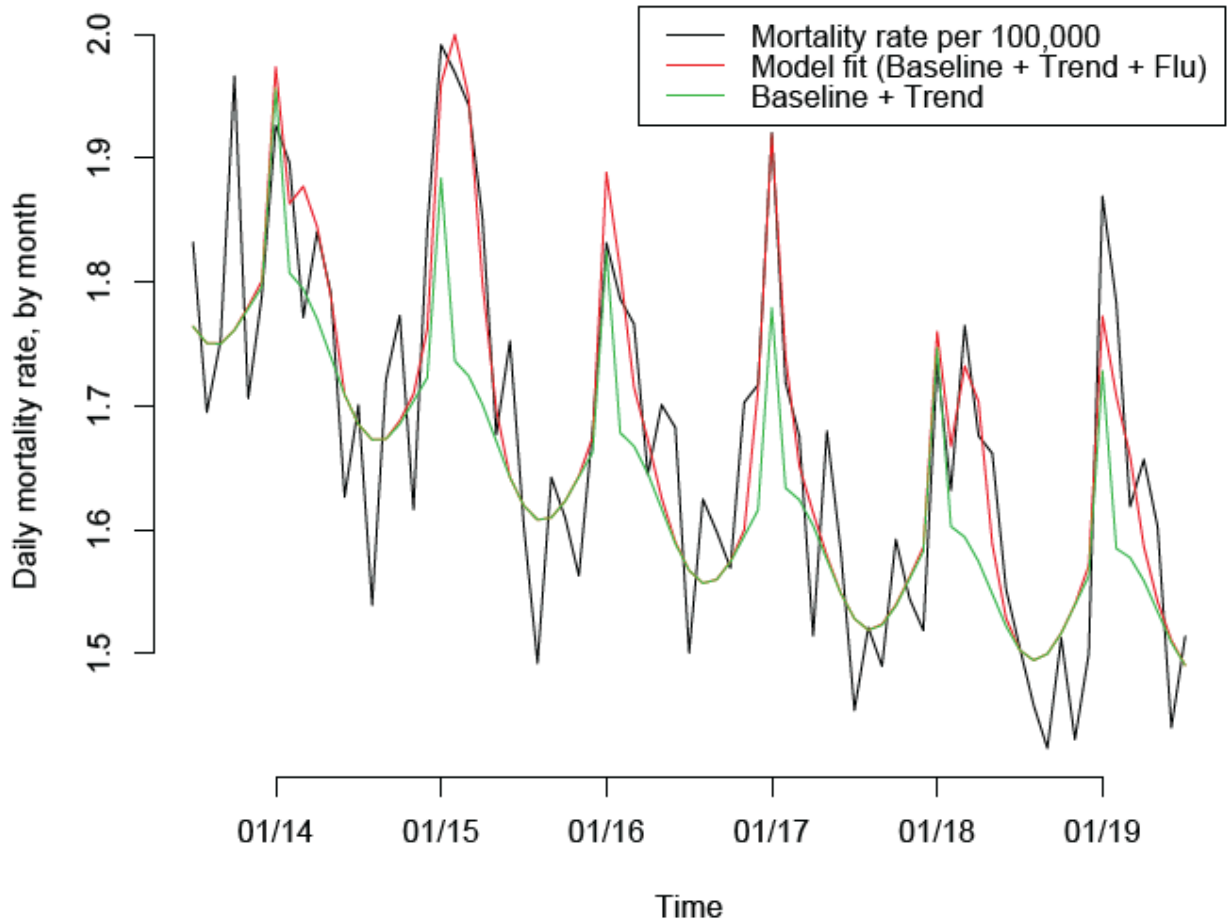

**Figure 4:** Model fit for the daily rates of circulatory deaths per 100,000 by month in Russia, 07/2013 though 07/2109. The contribution of influenza to circulatory mortality is represented by the difference between the red and the green curves.

Table 1 presents the estimates of the annual contribution of influenza to circulatory and respiratory mortality for each of the 2013/14 through the 2018/19 seasons (with each influenza season assumed to run from September through June), as well as the annual averages for the corresponding contributions for that time period (6 seasons). We estimate an average annual 17636 (95% CI (9482,25790)) deaths for circulatory causes and 4179 (3250,5109) deaths for respiratory causes during our study period. The largest number of both circulatory and respiratory deaths was estimated during the 2014/15 seasons when drift variants of both influenza A/H3N2 and influenza *B/Yamagata* circulated in Russia [18]. We note that this season was also associated with high rates of mortality in certain other countries, e.g. [7]. We also note that the levels of influenza vaccination in Russia rose significantly starting the 2016/17 season. Compared to the 2013/14 through the 2015/16 seasons (first 3 seasons), during the 2016/17 through the 2018/19 seasons, the volume of influenza-associated mortality declined by about 16.1%, or 3809 annual deaths. Finally, we note that the estimated ratio of circulatory to respiratory influenza-associated deaths in Russia (about 4.22-to-1, Table 1) is quite higher that the corresponding estimate (about 1.35-to-1, [3]) for the US. At the same time, the overall ratio of circulatory-to-respiratory deaths in Russia (about 14-to-1 in 2018, [20]) is much higher than the corresponding ratio in the US (about 3.08-to-1 in 2017, [25]).

| Season | Circulatory causes | Respiratory causes |
|---|---|---|
| 2013/14 | 12364 (5497,19230) | 3116 (2329,3902) |
| 2014/15 | 32298 (18071,46525) | 6689 (5019,8359) |
| 2015/16 | 13128 (1046,25211) | 3565 (2171,4959) |
| 2016/17 | 17462 (4376,30548) | 3540 (2045,5035) |
| 2017/18 | 17463 (6294,28632) | 4887 (3606,6167) |
| 2018/19 | 13101 (4760,21442) | 3280 (2330,4229) |
| Annual average | 17636 (9482,25790) | 4179 (3250,5109) |

**Table 1:** Annual and average annual number of influenza-associated circulatory and respiratory deaths (with 95% confidence intervals) for the 2013/14 through the 2018/19 influenza seasons in Russia.

Table 2 estimates the average annual number of circulatory and respiratory deaths for the 2013/14 through the 2018/19 seasons associated with the circulation of influenza A/H2N2, A/H1N1 and B. We estimate that during our study period, influenza A/H3N2 was responsible for 51.8% of all circulatory influenza-associated deaths and 37.2% of all respiratory influenza-associated deaths (Table 2 vs. Table 1); influenza A/H1N1 was responsible for 23.4% of all circulatory influenza-associated deaths and 29.5% of all respiratory influenza-associated deaths; influenza B was responsible for 24.9% of all circulatory influenza-associated deaths and 33.3% of all respiratory influenza-associated deaths.

| Influenza (sub)type | Circulatory deaths | Respiratory deaths |
|---|---|---|
| A/H3N2 | 9140 (2608,15672) | 1553 (805,2301) |
| A/H1N1 | 4100 (-156,8355) | 1233 (746,1721) |
| B | 4396 (-760,9553) | 1393 (799,1987) |

**Table 2:** Average annual number of circulatory and respiratory deaths associated with the circulation of influenza A/H3N2, A/H1N1 and B during the 2013/14 through the 2018/19 seasons.

**Discussion**

Influenza circulation is known to be associated with a substantial mortality burden in the Northern Hemisphere, e.g. [3-9]. At the same time, there is limited information on the volume of influenza-related mortality in Russia, with estimates based on deaths with influenza in the diagnosis [2,12-15] representing only a small fraction of all influenza-associated deaths and not allowing for the evaluation of the relative contribution of different influenza subtypes to the full volume of influenza-associated mortality (Introduction). In this paper, we applied the previously developed methodology that was used to evaluate the rates of influenza-associated mortality in several countries, e.g. [3,4,6-9] to estimate the rates of influenza-associated mortality for circulatory causes and respiratory causes in Russia during the 2013/14 through the 2018/19 influenza seasons. We estimate an average annual 17636 (95% CI (9482,25790)) circulatory death and 4179 (3250,5109) respiratory deaths during our study period. We note that the high rates of circulatory influenza-associated deaths are consistent with the high burden of circulatory mortality in Russia [20], as well as the evidence about the cardiovascular manifestations of influenza infection [10,11]. Influenza A/H3N2 was the largest contributor to each cause of death, with influenza A/H1N1 and B (mostly *B/Yamagata*, which is consistent with some evidence from other countries [26,21,22]) also making a sizeable contribution to the volume of influenza-associated mortality. Additionally, the significant increase in the influenza vaccination levels in Russia starting the 2016/17 season was associated with some reduction (16.1% on average) in the annual volume of influenza-associated mortality. We hope that this work may help inform mitigation efforts. In particular, our findings support further increases in vaccination coverage, particularly for individuals with circulatory disease, with the contribution of influenza *B/Yamagata* to mortality (particularly during seasons for which there was a mismatch between the circulating influenza B viruses and the virus recommended for the trivalent vaccine [18]) also supporting the use of quadrivalent vaccines. We also note in that regard that the 2018/19 influenza season saw minimal circulation of influenza B in Russia [17], raising the possibility of a major influenza B epidemic during the 2019/20 season; as influenza *B/Yamagata* strains are not present in the trivalent vaccine recommended for the 2019/20 season [27], this further supports the need for a quadrivalent influenza vaccine, particularly for individuals with cardiac disease. Additionally, administration of antiviral medications for certain population subgroups defined by age and presence of underlying health conditions (particularly circulatory diseases) may help mitigate the burden of severe outcomes during influenza epidemics in Russia.

Our paper has some limitations. We only had access to monthly mortality data; moreover, those data are operational, with some delays in reporting, and some unreported deaths during a given calendar year being reported for January of the next year [23]. Finer mortality data stratified by week/age group are needed to get a more comprehensive understanding of the contribution of influenza to mortality for different causes of death in Russia. Additionally, due to the relatively small number of data points in the regression framework adopted in this paper, we used a simple trigonometric model for the baseline rates of mortality not associated with influenza circulation. Previous work (e.g. Supporting

Information for [3]) suggests limited sensitivity of the estimates of the rates of influenza-associated mortality on the model for the baseline rates of non-influenza mortality.

We believe that despite those limitations, this work provides a national study of influenza-related mortality in Russia, giving evidence for the substantial burden of influenza-associated mortality, particularly for circulatory causes, delineating the contribution of different influenza (sub)types to mortality and suggesting likely effects of the recent increases in vaccination coverage on the volume of influenza-associated mortality. In particular, our work supports further increases in vaccination coverage levels, particularly for individuals with circulatory disease, as well as the use of quadrivalent influenza vaccines and of antiviral medications for certain population groups. We hope that this work would stimulate further efforts involving more granular data (in particular, weekly mortality data, as well as data stratified by age) to better understand the effect of influenza epidemics in Russia on mortality and help inform mitigation efforts for different population subgroups.